%% file: main_PRL_resubmit_2.tex
\input{preamble.tex}

\begin{document}

\preprint{APS/123-QED}

\title{Higher Josephson harmonics in a tunable double-junction transmon qubit}

\author{Ksenia Shagalov}
\email{ksenia.shagalov@nbi.ku.dk}
\affiliation{Center for Quantum Devices, Niels Bohr Institute, University of Copenhagen, 2100 Copenhagen, Denmark}
\affiliation{NNF Quantum Computing Programme, Niels Bohr Institute, University of Copenhagen, 2100 Copenhagen, Denmark}

\author{David Feldstein-Bofill}
\affiliation{Center for Quantum Devices, Niels Bohr Institute, University of Copenhagen, 2100 Copenhagen, Denmark}
\affiliation{NNF Quantum Computing Programme, Niels Bohr Institute, University of Copenhagen, 2100 Copenhagen, Denmark}

\author{Leo Uhre Jakobsen}
\affiliation{Center for Quantum Devices, Niels Bohr Institute, University of Copenhagen, 2100 Copenhagen, Denmark}
\affiliation{NNF Quantum Computing Programme, Niels Bohr Institute, University of Copenhagen, 2100 Copenhagen, Denmark}

\author{Zhenhai Sun}
\affiliation{Center for Quantum Devices, Niels Bohr Institute, University of Copenhagen, 2100 Copenhagen, Denmark}
\affiliation{NNF Quantum Computing Programme, Niels Bohr Institute, University of Copenhagen, 2100 Copenhagen, Denmark}

\author{Casper Wied}
\affiliation{Center for Quantum Devices, Niels Bohr Institute, University of Copenhagen, 2100 Copenhagen, Denmark}
\affiliation{NNF Quantum Computing Programme, Niels Bohr Institute, University of Copenhagen, 2100 Copenhagen, Denmark}

\author{Amalie T. J. Paulsen}
\affiliation{Center for Quantum Devices, Niels Bohr Institute, University of Copenhagen, 2100 Copenhagen, Denmark}
\affiliation{NNF Quantum Computing Programme, Niels Bohr Institute, University of Copenhagen, 2100 Copenhagen, Denmark}

\author{Johann Bock Severin}
\affiliation{Center for Quantum Devices, Niels Bohr Institute, University of Copenhagen, 2100 Copenhagen, Denmark}
\affiliation{NNF Quantum Computing Programme, Niels Bohr Institute, University of Copenhagen, 2100 Copenhagen, Denmark}

\author{Malthe A. Marciniak}
\affiliation{Center for Quantum Devices, Niels Bohr Institute, University of Copenhagen, 2100 Copenhagen, Denmark}
\affiliation{NNF Quantum Computing Programme, Niels Bohr Institute, University of Copenhagen, 2100 Copenhagen, Denmark}

\author{Clinton A. Potts}
\affiliation{Center for Quantum Devices, Niels Bohr Institute, University of Copenhagen, 2100 Copenhagen, Denmark}
\affiliation{NNF Quantum Computing Programme, Niels Bohr Institute, University of Copenhagen, 2100 Copenhagen, Denmark}
\affiliation{Department of Electrical and Software Engineering, University of Calgary, 2500 University Drive NW, Calgary, Alberta T2N 1N4, Canada}

\author{Anders Kringhøj}
\affiliation{Center for Quantum Devices, Niels Bohr Institute, University of Copenhagen, 2100 Copenhagen, Denmark}
\affiliation{NNF Quantum Computing Programme, Niels Bohr Institute, University of Copenhagen, 2100 Copenhagen, Denmark}

\author{Jacob Hastrup}
\affiliation{Center for Quantum Devices, Niels Bohr Institute, University of Copenhagen, 2100 Copenhagen, Denmark}
\affiliation{NNF Quantum Computing Programme, Niels Bohr Institute, University of Copenhagen, 2100 Copenhagen, Denmark}

\author{Karsten Flensberg}
\affiliation{Center for Quantum Devices, Niels Bohr Institute, University of Copenhagen, 2100 Copenhagen, Denmark}

\author{Svend Krøjer}
\affiliation{Center for Quantum Devices, Niels Bohr Institute, University of Copenhagen, 2100 Copenhagen, Denmark}
\affiliation{NNF Quantum Computing Programme, Niels Bohr Institute, University of Copenhagen, 2100 Copenhagen, Denmark}

\author{Morten Kjaergaard}
\email{mkjaergaard@nbi.ku.dk}
\affiliation{Center for Quantum Devices, Niels Bohr Institute, University of Copenhagen, 2100 Copenhagen, Denmark}
\affiliation{NNF Quantum Computing Programme, Niels Bohr Institute, University of Copenhagen, 2100 Copenhagen, Denmark}

\date{\today}

\begin{abstract}
Tunable Josephson harmonics open new avenues for qubit design. We demonstrate a superconducting circuit element consisting of a tunnel junction in series with a SQUID loop, yielding a Josephson potential whose harmonic content is strongly tunable by magnetic flux. Through spectroscopy of the first four qubit transitions, together with an effective single-mode model renormalized by the internal mode, we resolve a second harmonic with an amplitude up to $\sim10\%$ of the fundamental. We identify a flux sweet spot where the ground-state dispersive shift vanishes, achieved by balancing the dispersive couplings to the internal and qubit modes. This highly tunable element provides a route toward protected qubits and customizable nonlinear microwave devices.
\end{abstract}

\maketitle

Superconductor–insulator–superconductor (SIS) Josephson junctions (JJs) form the nonlinear element in most superconducting circuits \cite{blais_rev}. The corresponding Josephson potential is conventionally modeled by a single harmonic, $U(\phi)=-E_J\cos\phi$, with $E_J$ the Josephson energy \cite{SIS_JJ}, which underpins key circuit properties, including anharmonicity and charge-noise sensitivity. However, recent experiments have shown that SIS junctions can exhibit higher harmonics, i.e. terms proportional to $\cos(k\phi)$ with $k>1$, arising from pinholes in the junction or from inductive contributions of the leads \cite{Willsch_2024,MIT_SIS_harmonics,qudit_d_12, ioan_pop_2_harmonics}.

Generalizing these observations, the potential of a superconducting Josephson element can be described through a Fourier series,
\begin{equation} \label{eq:har}
    U(\phi) = \sum^\infty_{k=1}U_k\cos(k\phi),
\end{equation}
where $U_k$ are the harmonic coefficients with the $k$'th term corresponding to transmission of $k$ pairs of Cooper-pairs across the junction~\cite{CPR_JJ}.
Controlling the ratio between the first and higher Josephson harmonics thus enables modifications of the potential energy landscape of a superconducting circuit. 
Importantly, higher harmonic engineering has enabled the design of protected qubits with suppressed sensitivity to charge or flux noise \cite{Bell_2014, Larsen_paper, protected_qubit_2021, protecte_qubit_2022, luca_rhombus, grid_states, zero_pi_andras, coherence_limit_cos2phi} and superconducting diodes \cite{diode_2022, diode_2023, diode_SNS, diode_manfra}. 

One approach toward generating higher Josephson harmonics uses superconductor–normal-metal–superconductor (SNS) junctions \cite{jc_SNS} and hybrid superconductor–semiconductor weak links \cite{S-Sm-S_JJ}.
In SNS elements, multiple Andreev reflections enable multi–Cooper-pair tunneling, producing a strongly non-sinusoidal potential \cite{nonsinusoidal_EPR} with a substantially higher-harmonic content \cite{SNS_CPR_theory, SNS_CPR_exp, Larsen_paper}.
Such junctions have been used to demonstrate gate-tunable qubits \cite{nanowire_gatemon, de_Lange_2015, gatemon_vlad}, $\Phi_0$-junctions \cite{phi_0_Josephson}, and diode-like behavior \cite{diode_2022, diode_2023, diode_SNS, diode_manfra}.
However, hybrid platforms often suffer from frequency instability and reduced coherence, potentially due to losses in the semiconducting region \cite{david_paper, zhen_paper}.

Another approach relies on placing two SIS junctions in series, a configuration previously employed for flux-noise suppression in tunable transmons \cite{Jose_paper,weak_TT_russia,three_mode_coupler_russia}, protected readout \cite{jane_qubit_readout, devoret_readout, junction_readout, arm_qubit}, and scalable junction fabrication \cite{trilayer_JJ}. Under suitable approximations, this element exhibits an effective energy–phase relation analogous to that of a single-channel SNS junction, resulting in a non-sinusoidal potential governed by the junction asymmetry \cite{Akhmerov_paper}.
Experiments conducted on InAs/Al hybrid heterostructures further demonstrate that tuning the junction asymmetry through gate tuning can modify the energy-phase relation from a sinusoidal form to one that closely resembles the SNS junction \cite{Luca_paper}.

In this manuscript, we introduce a tunable SIS-SIS circuit element comprising of a superconducting quantum interference device (SQUID) \cite{SQUID, engineers_guide} in series with a single JJ, enabling \textit{in situ} control of the energy-phase relation via an external magnetic flux, $\Phi_\mathrm{e}$, and suitable for integration into a broad range of superconducting circuits. We implement this element as a tunable double-junction transmon, shunted by a large capacitor and operated in the transmon regime.
Spectroscopy of higher-order qubit transitions shows an enhanced higher-harmonic contribution compared to previously reported values in SIS junctions \cite{Willsch_2024,MIT_SIS_harmonics,qudit_d_12, ioan_pop_2_harmonics}. 
A model including an auxiliary (internal) mode, associated with the superconducting island between the JJ and SQUID, captures the spectrum and accounts for a renormalization of the qubit frequency. We further observe dispersive coupling of this internal mode to the readout resonator, leading to a flux point where the total ground-state dispersive shift cancels.

\begin{figure}[t!]
\advance\leftskip-0.5cm
\includegraphics[width=\columnwidth]{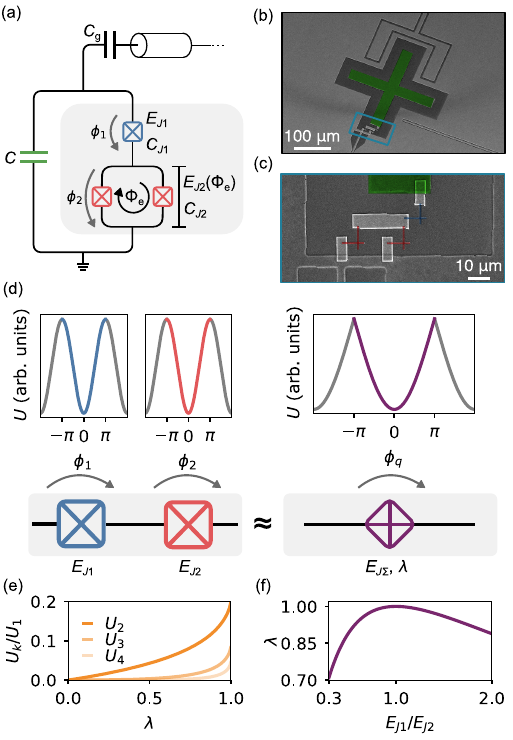} 
\caption{\label{fig_device} Qubit circuit and device design. (a) Circuit schematic of the tunable double-junction transmon with a single JJ (blue) and a SQUID loop (red) in series, shunted by a large capacitor (green). (b) SEM of the tunable double-junction transmon showing the large island which forms the capacitor (green). A quarter-wave resonator is capacitively coupled to the qubit for readout. A flux line threads magnetic flux through the SQUID loop, and the drive line facilitates the XY control. (c) SEM image of the single JJ (blue) and the SQUID loop (red) in series. (d) The potentials $U$ plotted for each SIS junction (blue, red), which in series effectively correspond to the potential from Eq.~\ref{eq_pot_Akhmerov}, shown for $\lambda = 1$ (purple). (e) Harmonic content of the potential from Eq.~\ref{eq_pot_Akhmerov} as a function of $\lambda$. (f) $\lambda$ as a function of the ratio between the Josephson energies of the two junctions.}
\centering
\end{figure}

Figure~\ref{fig_device}a shows a circuit diagram of the tunable double-junction transmon, with the scanning electron microscope (SEM) images of the corresponding fabricated device shown in Figs.~\ref{fig_device}b, c (fabrication details can be found in Supplementary Material~\cite{SM}). We denote by $\phi_1$ and $\phi_2$ the superconducting phases across the single junction and the SQUID, respectively. To understand the origin of the appreciable higher harmonics in an SIS–SIS device, we consider a ``reduced'' model in which the charging energy of the middle island is initially omitted. 
In this limit, the two junctions behave as a single effective element with a potential of the form \cite{Akhmerov_paper}
\begin{equation} \label{eq_pot_Akhmerov}
U^{\textrm{red.}}_{\jj}(\phi_q) = -E_{J\Sigma}\sqrt{1 - \lambda\sin^2(\phi_q/2)},
\end{equation}
where $\phi_q = \phi_1 + \phi_2$ is the phase of the qubit mode, $E_{J\Sigma}=E_{J1}+E_{J2}$, and
\begin{equation}
\lambda = 4E_{J1}E_{J2} / {(E_{J1}+E_{J2})^2}
\label{eq_asymmetry}
\end{equation}
is the junction asymmetry parameter.
The resulting potential landscape in this reduced model is shown in Fig.~\ref{fig_device}d.
The emergence of higher harmonics becomes explicit in the symmetric case $E_{J1}=E_{J2}$, where $\lambda=1$ and Eq.~\eqref{eq_pot_Akhmerov} expands to
\begin{equation}
U^\textrm{red.}_{\jj}(\phi_q)
= E_{J\Sigma}\sum_k \left[(-1)^{k+1}\frac{4}{\pi - 4\pi k^2}\right]\cos(k\phi_q),
\end{equation}
identifying the coefficients in brackets with $U_k$ from Eq.~\eqref{eq:har}.
Figure~\ref{fig_device}e shows that the second harmonic can reach $|U_2/U_1|\approx 0.2$ at $\lambda=1$~\cite{leo_paper}, substantially larger than previously reported values for single SIS junctions \cite{Willsch_2024,MIT_SIS_harmonics,qudit_d_12, ioan_pop_2_harmonics}.
Conversely, for $\lambda\ll 1$, the familiar sinusoidal form is recovered: $ U^\textrm{red.}_{\jj}(\phi_q)\propto \cos\phi_q$.
Figure~\ref{fig_device}f shows how $\lambda$ varies with $E_{J1}/E_{J2}$, reflecting its role as the junction asymmetry parameter, analogous to the effective transmission $\tau$ of a single Andreev mode in the SNS potential $U_\mathrm{SNS}(\phi)=-\Delta\sqrt{1-\tau\sin^2(\phi/2)}$ where $\Delta$ is the superconducting gap~\cite{Beenakker}.

In the tunable double-junction transmon, threading the external flux through the SQUID loop makes the junction asymmetry parameter flux dependent, $\lambda \rightarrow \lambda(\Phi_\mathrm{e})$, and thereby allows \textit{in situ} control of the effective potential and its harmonic content.
As Fig.~\ref{fig_device}f shows, $\lambda$ varies slowly near $E_{J1}/E_{J2}\approx 1$, so precise flux biasing to the symmetric point is not required for higher harmonics to be appreciable.
In practice, whenever $E_{J2}(\Phi_\mathrm{e})$ lies within roughly a factor of two of $E_{J1}$, the asymmetry parameter remains above $\lambda \approx 0.9$, corresponding to a second-harmonic ratio of $|U_2/U_1|\approx 0.12$.

\begin{figure}[t!]
\advance\leftskip-0.5cm
\includegraphics[width=\columnwidth]{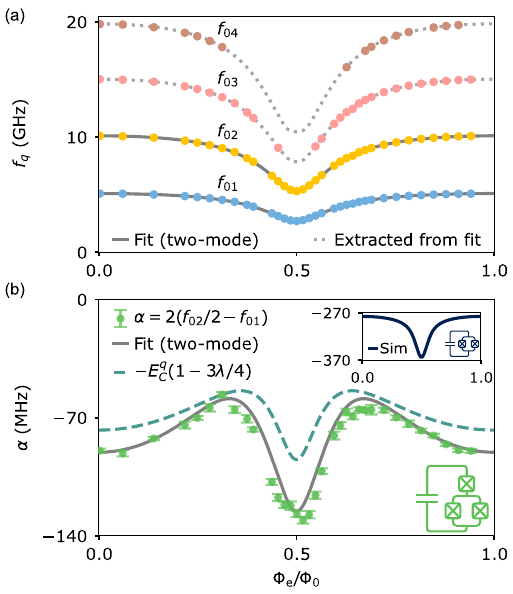}
\caption{\label{fig_spec} Flux-dependent two-tone spectroscopy of the tunable double-junction transmon (second cooldown) and fit using the two-mode model from Eq.~\ref{eq: full_ham}. (a) Two-tone spectroscopy of the $f_{01}$ (light blue), $f_{02}$ (yellow), $f_{03}$ (pink) and $f_{04}$ (brown) transitions as a function of $\Phi_\mathrm{e}$. The $f_{01}$ and $f_{02}$ transitions are fitted to the two-mode model of Eq.~\ref{eq: full_ham} (solid gray), and the resulting parameters are used to predict $f_{03}$ and $f_{04}$ (dotted gray). (b) Extracted anharmonicity (green) obtained from the $f_{01}$ and $f_{02}/2$ transitions, together with the corresponding fit (solid gray). Error bars indicate the statistical uncertainty extracted from the Lorentzian fits. The analytical expression from Eq.~\ref{eq_anh}, based on the potential in Eq.~\ref{eq_pot_Akhmerov}, is shown for comparison (dashed teal). Inset: simulated anharmonicity versus flux for a tunable transmon with comparable device parameters.}
\centering
\end{figure}

In Fig.~\ref{fig_spec}a we show the first four transitions of the tunable double-junction transmon as a function of flux threading the SQUID loop. 
We observe through two-tone spectroscopy the single-photon $f_{01}$ transition together with the multi-photon transitions $f_{02}/2$, $f_{03}/3$, and $f_{04}/4$, from which the higher transition frequencies were extracted through Lorentzian fitting. To fit the spectra as a function of external flux, we use a two-mode Hamiltonian that includes both phase degrees of freedom across the single junction and the SQUID,
\begin{equation}\label{eq: full_ham}
    H_{\textrm{full}} = 4E_{C1}n_1^2 + 4E_{C2}n_2^2 + g_{12}n_1n_2 + U_{\jj}^\mathrm{full}(\phi_1,\phi_2),
\end{equation}
with 
\begin{equation}
U_{\jj}^\mathrm{full}(\phi_1,\phi_2) = -E_{J1}\cos(\phi_1) - E_{J2}\cos(\phi_2),
\label{eq:two-mode}
\end{equation}
where $E_{C1}, E_{C2}$, is the charging energy of each of the modes, $n_{1}, n_{2}$ their corresponding charge operators and $g_{12}$ the coupling between them. We perform a simultaneous fit to all measured $f_{01}$ and $f_{02}/2$ values, extracting the island capacitance as well as the capacitances and Josephson energies of the individual junctions.
The resulting fit (solid line in Fig.~\ref{fig_spec}a) also accurately predicts $f_{03}$ and $f_{04}$ (dotted gray lines), demonstrating quantitative agreement with the measured spectrum. Details of the circuit model and fitting procedure are provided in Supplementary Material~\cite{SM}.

In Fig.~\ref{fig_spec}b, we extract the flux-dependent anharmonicity via,
$\alpha = 2\!\left(f_{02} / 2 - f_{01}\right)$, shown as green circles. The solid line is the anharmonicity obtained from the fit in Fig.~\ref{fig_spec}a, again demonstrating good agreement between the two-mode model and the measured spectra. The scatter of the anharmonicity data, on the order of several $\si{\mega\hertz}$, is consistent with the uncertainty from Lorentzian fits and may reflect temporal frequency fluctuations typical of superconducting devices.
Despite similar configurations being previously employed as weakly flux-tunable transmons \cite{Jose_paper,weak_TT_russia,three_mode_coupler_russia}, the tunable double-junction transmon exhibits a pronounced non-monotonic flux dependence of the anharmonicity between $\Phi_\mathrm{e}=0$ and $\Phi_\mathrm{e}=0.5\Phi_0$. This behavior differs fundamentally from a conventional transmon, where the anharmonicity varies monotonically with flux. The inset of Fig.~\ref{fig_spec}b illustrates this comparison using a standard transmon model with identical charging energy, $E_C$, and matched frequency tunability. Notably, at $\Phi_\mathrm{e}=0$ the tunable double-junction transmon reaches $h\alpha \approx -E_C/3$ (Fig.~\ref{fig_spec}b), deviating significantly from the typical transmon value $h\alpha \approx -E_C$~\cite{transmon}.

The qualitative difference between the anharmonicity of the tunable double–junction transmon and that of a conventional transmon can be understood from the reduced model of Eq.~\eqref{eq_pot_Akhmerov}.  
Using the reduced potential, the Hamiltonian of the qubit mode is
\begin{equation} \label{eq:effective_ham}
    H_\mathrm{red.} = 4E_C^q n_q^2 + U^{\textrm{red.}}_{\jj}(\phi_q),
\end{equation}
where $E_C^q = e^2/[2(C + C_{J1}C_{J2}/(C_{J1}+C_{J2}))]$ is the charging energy of the qubit mode.
Applying the transmon approximation ($E^\textrm{eff}_{J}/E_C^q \gg 1$) and expanding $U^\textrm{red.}_{\jj}(\phi_q)$ to fourth order yields 
$hf_{01} \approx \sqrt{8E_C^q E_J^\mathrm{eff}} - h\alpha$, with 
$E_J^\mathrm{eff}=E_{J1}E_{J2}/(E_{J1}+E_{J2})$,  
and an anharmonicity
\begin{equation} \label{eq_anh}
    h\alpha \approx -E_C^q\!\left(1-\frac{3}{4}\lambda\right),
\end{equation}
result previously demonstrated from studying anharmonicity in gatemons~\cite{anharmonicity_paper}.
Figure~\ref{fig_spec}b shows Eq.~\eqref{eq_anh} (teal dashed line) evaluated using parameters extracted from the two-mode fit.  
The remaining discrepancy between Eq.~\eqref{eq_anh} and the measured anharmonicity arises from the limitations of the transmon approximation and from the internal mode, which renormalizes the qubit spectrum.

We now investigate the presence of higher harmonics in the tunable double-junction transmon.
In the two-mode model (Eq.~\eqref{eq: full_ham}), each junction is described by a sinusoidal potential; thus,  higher harmonic content of the qubit mode $\phi_q$ is not well-defined. It emerges only after eliminating the internal degree of freedom and projecting onto an effective single-mode picture, analogous to the emergence of an effective linear inductance in junction-array superinductors~\cite{Manucharyan2009,DiPaolo2021}. The reduced model (Eq.~\eqref{eq:effective_ham}) captures this emergence qualitatively, but does not accurately reproduce the measured spectra (see Supplementary Material~\cite{SM}). To reliably assess the higher harmonic content, we therefore derive an approximate single-mode model that incorporates the leading-order effect of the internal mode.

In the situation where the excitations in the internal mode are higher in frequency than the first few qubit excitations, we can apply a Born-Oppenheimer (BO) approximation to describe the low-energy spectrum of the qubit~\cite{terhal_lecture_notes, Rymarz2023}. In this approximation, the ground-state energy of the internal mode depends parametrically on $\phi_q$ and therefore acts as an effective potential for the qubit mode.
Within an analytical BO approximation framework, we can utilize that the internal mode is transmon-like, i.e. $E^\textrm{int}_C / E_{J\Sigma} \ll 1$, resulting in a BO correction due to the internal mode, up to first order in $E^\textrm{int}_C / E_{J\Sigma}$, given by \cite{leo_paper},
\begin{align} \label{eq_pot_BO}
        U^\text{BO}_{\jj}({\phi}_q) =  E_{J\Sigma}\sqrt{\frac{2E^\textrm{int}_{C}}{E_{J\Sigma}} \sqrt{1 - \lambda \sin^2\!\left({\phi}_q/2\right)}},
\end{align}
where $E^{\textrm{int}}_C = e^2 / [2(C_{J1} + C_{J2})]$ is the charging energy of the internal mode and $E^\textrm{int}_C / E_{J\Sigma}\sim1/300$ for our experimental parameters.
Thus, our one-dimensional BO model for the qubit mode is given by
\begin{equation}
    H_\textrm{BO} = 4E^q_Cn^2_q + U^{\mathrm{red.}}_{\jj}(\phi_q) + U^\text{BO}_{\jj}(\phi_q).
\end{equation}
The harmonic content of this BO model can be extracted via
\begin{equation} \label{eq:U_eff_harm}
    U_k = \frac{1}{\pi}\int_{-\pi}^\pi \left[ U^\mathrm{red.}_{\jj}(\phi_q) + U^\text{BO}_{\jj}(\phi_q)\right]\cos(k\phi_q)d\phi_q.
\end{equation}

\begin{figure}[t!]
\advance\leftskip-0.5cm
\includegraphics[width=\columnwidth]{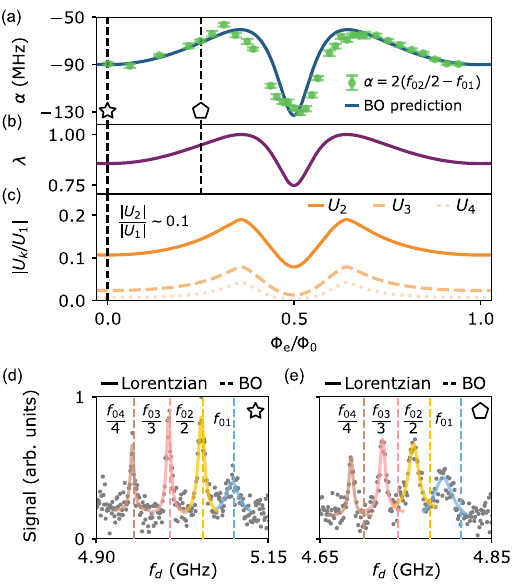}
\caption{\label{fig_harmonics}
Measured anharmonicity as a function of flux and extracted harmonic content of the tunable double-junction transmon (second cooldown).
(a) Measured anharmonicity values with error bars (green) compared with the Born–Oppenheimer (BO) prediction (blue) using the fit parameters extracted from the two-mode model. Vertical dashed lines mark $\Phi_\mathrm{e}=0$ and $\Phi_\mathrm{e}=0.25\Phi_0$.
(b) Extracted junction asymmetry parameter, $\lambda$. Vertical dashed lines mark $\lambda = 0.85$ and $\lambda = 0.95$. (c) Extracted second, third, and fourth harmonics normalized to the first harmonic from the BO model; at $\Phi_\mathrm{e}=0$ we obtain $U_k = [1,0.107,0.023,0.006]$.
(d) Four observed transitions at $\Phi_\mathrm{e}=0$ with Lorentzian fits (solid) and BO-predicted frequencies (dashed), showing good agreement.
(e) Four observed transitions at $\Phi_\mathrm{e}=0.25\Phi_0$; here the BO model captures the relative level spacings (anharmonicity) but not the absolute transition frequencies.}
\centering
\end{figure}

In Fig.~\ref{fig_harmonics}a, we compare the anharmonicity predicted by the BO model (blue solid line) using the fit parameters extracted from the two-mode model, with the measured values (green points) and find excellent agreement. We extract the junction asymmetry parameter, $\lambda$, across flux (Fig.~\ref{fig_harmonics}b) which determines the magnitude of the higher harmonics (Fig.~\ref{fig_harmonics}c). While junction asymmetry in the SQUID reduces the accessible tuning range of $\lambda$, it does not qualitatively alter the emergence of higher harmonics. Fig.~\ref{fig_harmonics}d shows a representative spectroscopy measurement at $\Phi_\mathrm{e} = 0$, averaged over drive amplitudes, resulting in power broadening of the $f_{01}$ transition.
The data are fitted with Lorentzians (solid lines) and compared to BO predictions based on the two-mode model fit parameters (dashed lines), again showing excellent agreement. Using Eq.~\eqref{eq:U_eff_harm} we extract the first four harmonic coefficients $U_k = [1, 0.107, 0.023, 0.006]$ at $\lambda = 0.85$, in quantitative agreement with Fig.~\ref{fig_device}e and significantly exceeding values reported for single SIS junctions \cite{Willsch_2024,MIT_SIS_harmonics,qudit_d_12, ioan_pop_2_harmonics}.

Despite accurately capturing the anharmonicity across all flux values, the BO model fails to reproduce the transition frequencies at $\Phi_\mathrm{e}=0.25\Phi_0$ (Fig.~\ref{fig_harmonics}e, dashed lines), due to the  strong asymmetry between junction energies ($E_{J2}<E_{J1}$) and capacitances ($C_{J1}<C_{J2}$) around $\Phi_\mathrm{e}=0.44\Phi_0$. In contrast, the two-mode model remains valid and accurately reproduces the transition frequencies. Extending it to include dispersive coupling to the resonator further provides access to the internal-mode dynamics (see Supplementary Material~\cite{SM}).

\begin{figure}[t!]
\advance\leftskip-0.5cm
\includegraphics[width=\columnwidth]{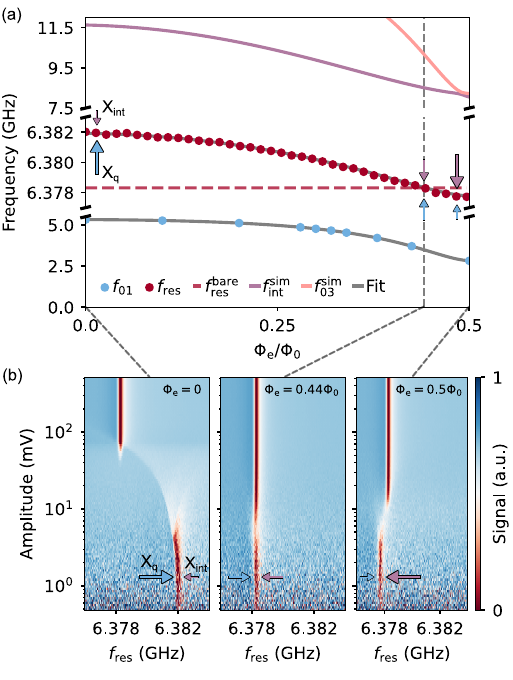}
\caption{\label{fig_int} Flux dependence of the total dispersive shift in the tunable double-junction transmon (first cooldown). (a) 
Measured data points of the qubit $f_{01}$ (light blue) and of the resonator $f_\textrm{res}$ (dark red) with the bare frequency of the resonator $f^\textrm{bare}_\textrm{res}$ indicated by the red dashed line. Extending the two-mode model from Eq.~\ref{eq: full_ham} (see Supplementary Material~\cite{SM}), we fit $f_\textrm{res}$ (gray) and extract the simulated internal-mode frequency $f^\mathrm{sim}_\mathrm{int}$ (purple) and the qubit’s $f^\mathrm{sim}_{03}$ transition, revealing an avoided crossing near $\Phi_\mathrm{e}=0.5\Phi_0$. (b) Resonator spectroscopy versus readout amplitude at three representative flux points. At $\Phi_\mathrm{e}=0$ (left), the qubit contribution dominates, shifting the resonator upward. At the balance point (middle), the qubit and internal-mode contributions cancel, giving zero net dispersive shift. Near $\Phi_\mathrm{e}=0.5\Phi_0$ (right), the internal mode dominates, shifting the resonator downward.}
\centering
\end{figure}

We measure the resonator frequency with the qubit in its ground state as a function of flux (red points in Fig.~\ref{fig_int}a), $f_\mathrm{res}$, and fit it simultaneously with $f_{01}$ using the extended two-mode model, thereby extracting the internal-mode frequency (purple) and the coupling to the resonator.  
Since the qubit transition remains below the bare resonator frequency at $6.378~\si{\giga\hertz}$ (red dashed line), the qubit mode alone would produce a positive dispersive shift. In the tunable double-junction transmon, however, the measured ground-state dispersive shift, $\chi_0$, arises from the combined contributions of the qubit and the internal mode, $\chi_0 = \chi_q + \chi_{\mathrm{int}}$. At $\Phi_\mathrm{e}=0$, $\chi_q$ dominates, yielding an upward shift (Fig.~\ref{fig_int}b, left); at $\Phi_\mathrm{e}=0.44\Phi_0$, the contributions cancel, giving $\chi_0=0$ (middle); and near $\Phi_\mathrm{e}=0.5\Phi_0$, the internal mode dominates, producing a downward shift (right).
The flux tunability of $\chi_0$, and in particular the observation of $\chi_0=0$, motivates exploration of regimes where the qubit-state-dependent dispersive shift, $\chi_{01}=\chi_1-\chi_0$, is suppressed, which may be exploited for next-generation qubit readout, particularly when combined with a flux sweet-spot~\cite{cloaking_blais, cloaking_blais_2,cloaking_wallraff,fluxonium_readout_theory,fluxonium_readout_exp, pmon}.

In summary, we have realized and characterized a tunable double-junction transmon composed of a SQUID in series with a single SIS junction, enabling \textit{in situ} flux control of the energy-phase relation. We identified flux regimes where the circuit behaves as a single effective mode and extracted substantial higher-harmonic content, including a second harmonic exceeding previously reported values for individual SIS junctions. 
The internal mode associated with the middle-island charging energy both renormalizes the qubit spectrum, captured in part by a Born–Oppenheimer correction, and couples dispersively to the readout resonator. Due to its large energy separation from the qubit mode and the absence of a direct drive matrix element coupling the two modes, leakage is suppressed. Moreover, as the internal mode resides in the transmon regime, charge dispersion is reduced and it is therefore not expected to introduce significant additional decoherence. Flux-dependent resonator spectroscopy further reveals a balance point where the ground-state dispersive shifts from the qubit and internal modes cancel, yielding a vanishing net shift. These results establish the tunable SIS–SIS element as a building block for superconducting circuits, operating analogously to a conventional SQUID while providing access to higher harmonics within a fully superconducting framework. This enables routes toward protected-qubit designs such as the $\cos(2\phi)$ qubit, which require suppression of the fundamental harmonic, e.g., via interference in a SQUID geometry using multiple SIS–SIS elements \cite{natasha_paper} or hybrid combinations with SNS junctions \cite{david_qppq}.

\begin{acknowledgments}
We gratefully acknowledge useful discussions with Valla Fatemi, Christian Andersen, Nataliia Zhurbina and Tyra Cortinez Samenius. 

This research was supported by the Novo Nordisk Foundation (grant no. NNF22SA0081175), the NNF Quantum Computing Programme (NQCP), Villum Foundation through a Villum Young Investigator grant (grant no. 37467), the Innovation Fund Denmark (grant no. 2081-00013B, DanQ), the U.S. Army Research Office (grant no. W911NF-22-1-0042, NHyDTech), by the European Union through an ERC Starting Grant, (grant no. 101077479, NovADePro), and by the Carlsberg Foundation (grant no. CF21-0343). C.A.P. acknowledges the support of the Natural Sciences and Engineering Research Council of Canada (NSERC) (No. RGPIN-2026-05057). Any opinions, findings, conclusions or recommendations expressed in this material are those of the author(s) and do not necessarily reflect the views of Army Research Office or the US Government. 
Views and opinions expressed are those of the author(s) only and do not necessarily reflect those of the European Union or the European Research Council. Neither the European Union nor the granting authority can be held responsible for them. 
Finally, we gratefully acknowledge Lena Jacobsen for program management support.
\end{acknowledgments}

\section*{Data Availability}

The data that support the findings of this article are openly available \cite{data}.

\input{references.bbl}
\end{document}


\preprint{APS/123-QED}

\title{Supplementary Material: \\ Higher Josephson harmonics in a tunable double-junction transmon qubit}

\author{Ksenia Shagalov}
\email{ksenia.shagalov@nbi.ku.dk}
\affiliation{Center for Quantum Devices, Niels Bohr Institute, University of Copenhagen, 2100 Copenhagen, Denmark}
\affiliation{NNF Quantum Computing Programme, Niels Bohr Institute, University of Copenhagen, 2100 Copenhagen, Denmark}

\author{David Feldstein-Bofill}
\affiliation{Center for Quantum Devices, Niels Bohr Institute, University of Copenhagen, 2100 Copenhagen, Denmark}
\affiliation{NNF Quantum Computing Programme, Niels Bohr Institute, University of Copenhagen, 2100 Copenhagen, Denmark}

\author{Leo Uhre Jakobsen}
\affiliation{Center for Quantum Devices, Niels Bohr Institute, University of Copenhagen, 2100 Copenhagen, Denmark}
\affiliation{NNF Quantum Computing Programme, Niels Bohr Institute, University of Copenhagen, 2100 Copenhagen, Denmark}

\author{Zhenhai Sun}
\affiliation{Center for Quantum Devices, Niels Bohr Institute, University of Copenhagen, 2100 Copenhagen, Denmark}
\affiliation{NNF Quantum Computing Programme, Niels Bohr Institute, University of Copenhagen, 2100 Copenhagen, Denmark}

\author{Casper Wied}
\affiliation{Center for Quantum Devices, Niels Bohr Institute, University of Copenhagen, 2100 Copenhagen, Denmark}
\affiliation{NNF Quantum Computing Programme, Niels Bohr Institute, University of Copenhagen, 2100 Copenhagen, Denmark}

\author{Amalie T. J. Paulsen}
\affiliation{Center for Quantum Devices, Niels Bohr Institute, University of Copenhagen, 2100 Copenhagen, Denmark}
\affiliation{NNF Quantum Computing Programme, Niels Bohr Institute, University of Copenhagen, 2100 Copenhagen, Denmark}

\author{Johann Bock Severin}
\affiliation{Center for Quantum Devices, Niels Bohr Institute, University of Copenhagen, 2100 Copenhagen, Denmark}
\affiliation{NNF Quantum Computing Programme, Niels Bohr Institute, University of Copenhagen, 2100 Copenhagen, Denmark}

\author{Malthe A. Marciniak}
\affiliation{Center for Quantum Devices, Niels Bohr Institute, University of Copenhagen, 2100 Copenhagen, Denmark}
\affiliation{NNF Quantum Computing Programme, Niels Bohr Institute, University of Copenhagen, 2100 Copenhagen, Denmark}

\author{Clinton A. Potts}
\affiliation{Center for Quantum Devices, Niels Bohr Institute, University of Copenhagen, 2100 Copenhagen, Denmark}
\affiliation{NNF Quantum Computing Programme, Niels Bohr Institute, University of Copenhagen, 2100 Copenhagen, Denmark}
\affiliation{Department of Electrical and Software Engineering, University of Calgary, 2500 University Drive NW, Calgary, Alberta T2N 1N4, Canada}

\author{Anders Kringhøj}
\affiliation{Center for Quantum Devices, Niels Bohr Institute, University of Copenhagen, 2100 Copenhagen, Denmark}
\affiliation{NNF Quantum Computing Programme, Niels Bohr Institute, University of Copenhagen, 2100 Copenhagen, Denmark}

\author{Jacob Hastrup}
\affiliation{Center for Quantum Devices, Niels Bohr Institute, University of Copenhagen, 2100 Copenhagen, Denmark}
\affiliation{NNF Quantum Computing Programme, Niels Bohr Institute, University of Copenhagen, 2100 Copenhagen, Denmark}

\author{Karsten Flensberg}
\affiliation{Center for Quantum Devices, Niels Bohr Institute, University of Copenhagen, 2100 Copenhagen, Denmark}

\author{Svend Krøjer}
\affiliation{Center for Quantum Devices, Niels Bohr Institute, University of Copenhagen, 2100 Copenhagen, Denmark}
\affiliation{NNF Quantum Computing Programme, Niels Bohr Institute, University of Copenhagen, 2100 Copenhagen, Denmark}

\author{Morten Kjaergaard}
\email{mkjaergaard@nbi.ku.dk}
\affiliation{Center for Quantum Devices, Niels Bohr Institute, University of Copenhagen, 2100 Copenhagen, Denmark}
\affiliation{NNF Quantum Computing Programme, Niels Bohr Institute, University of Copenhagen, 2100 Copenhagen, Denmark}

\date{\today}

\maketitle

\renewcommand{\thesection}{}
\renewcommand{\thesubsection}{}
\makeatother

\setcounter{tocdepth}{2}
\tableofcontents

\renewcommand{\thefigure}{S\arabic{figure}}
\setcounter{figure}{0}

\renewcommand{\thetable}{S\arabic{table}}
\setcounter{table}{0}

\section{Fabrication and Measurement Setup}
The qubit island was patterned from a $200 \, \si{\nano\meter}$ thick Al film evaporated on a high-resistivity silicon substrate, with a capacitance to ground that corresponds to a charging energy of $E_C/h \approx 305 \, \si{\mega\hertz}$. The JJs were fabricated using the double-angle shadow evaporation technique with controlled \textit{in situ} oxidation at a high pressure of $120 \, \si{\milli\bar}$ for six minutes which forms a stack of Al/AlO\textsubscript{x}/Al, as illustrated in the scanning electron microscope (SEM) image in Figure 1 of the main text. Due to the thick oxide layer, the JJ area for a given $E_J$ was designed to be relatively large resulting in a large intrinsic junction capacitance, $C_{J1}, C_{J2}$. A patch layer was then deposited to ensure galvanic connection between the SQUID loop, the single JJ, and the rest of the circuit, following the removal of the native aluminum oxide using \textit{in situ} argon milling. Nearby transmission lines facilitate qubit control, with the shorted transmission line regulating the external magnetic flux by adjusting the current, while the open transmission line drives microwave excitations. The qubit is read out using a quarter-wave resonator with a resonance frequency of $f_\textrm{res} = 6.378 \, \si{\giga\hertz}$, and a coupling capacitance to the qubit $C_g = 7.3 \, \si{\femto\farad}$. 

\begin{table*}[b]
\begin{ruledtabular}
\begin{tabular}{ccccccccccc}
 &$C \, (\si{\femto\farad})$&$E_{J1} \,  (h\si{\giga\hertz})$&$C_{J1} \,(\si{\femto\farad})$&$E_{JA} \, (h\si{\giga\hertz})$&$C_{JA} \, (\si{\femto\farad})$&$E_{JB} \, (h\si{\giga\hertz})$&$C_{JB} \, (\si{\femto\farad})$ & $f^{\textrm{bare}}_\textrm{res} (\si{\giga\hertz})$ & $C_{g} (\si{\femto\farad})$ & $C_{r} (\si{\pico\farad})$ \\ \hline
 CD1 & 63.3 & 25.7 & 27.8 & 31.9 & 34.5 & 23.6 & 25.6 & 6.3783 & 7.3 & 1.2\\
 CD2 & 63.3 & 23.4 & 27.8 & 30.0 & 34.5 & 22.2 & 25.6 \\
\end{tabular}
\end{ruledtabular}
\caption{\label{tab:parameter_tab}
Device parameters in the first cooldown (CD1) and the second cooldown (CD2) extracted from fitting of the transition frequencies as a function of flux to the two-mode model. 1) Qubit island capacitance. 2) Josephson energy of the single junction. 3) Capacitance of the single junction. 4) Josephson energy of the left junction in the SQUID loop. 5) Capacitance of the left junction in the SQUID loop. 6) Josephson energy of the right junction in the SQUID loop. 7) Capacitance of the right junction in the SQUID loop. 8) Bare resonator frequency. 9) Qubit-resonator coupling capacitance. 9) Resonator self-capacitance.}
\end{table*}

\begin{figure}[t]
\advance\leftskip-0.5cm
\includegraphics[width=\columnwidth]{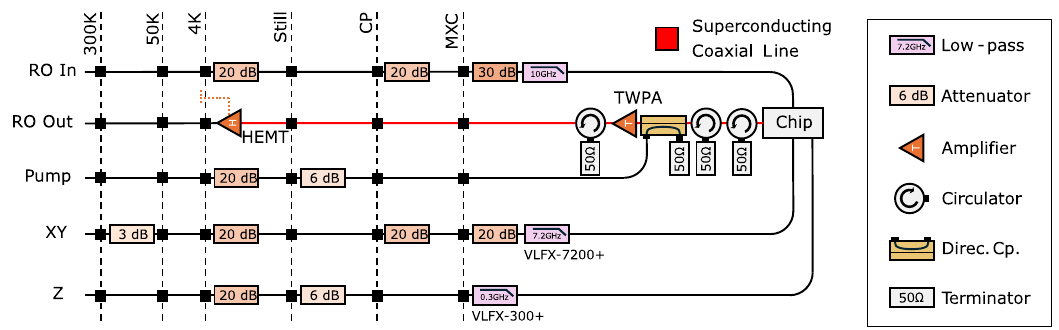} 
\caption{\label{fig_fridge_diagram} Wiring schematic of the experimental setup.}
\centering
\end{figure}

Measurements were performed in a dilution refrigerator (see wiring in Fig.~\ref{fig_fridge_diagram}) with a base temperature of $\sim8 \, \si{\milli\kelvin}$, over two independent cooldowns of the same device. The control hardware consists of a Quantum Machines OPX1000 chassis containing front-end modules (FEMs) for controlling both the XY-control and readout microwave signals (MW-FEM) and low frequency (LF-FEM) for baseband flux control for the Z-control lines. All experiments were orchestrated with an in-house python data acquisition platform called \textit{Pelagic}. Data from the first cooldown (CD1) is shown in Figure 4 in the main text, and Fig.~\ref{fig_supp_coherence} and data from the second cooldown (CD2) is shown in Figures 2, 3 in the main text and Figs.~\ref{fig_supp_analysis},~\ref{fig_supp_full_fit} ,~\ref{fig_supp_FFT}. Table~\ref{tab:parameter_tab} summarizes the parameters that were extracted from the fitting of the spectroscopy measurements as a function of flux to the two-mode model. In the first cooldown (CD1) the parameters were fit to the device coupled to the resonator while fitting $f_{01}, f_{02}/2$ and $f_\textrm{res}$ in parallel. In the second cooldown (CD2) we used the capacitances found in the first cooldown and fit only the Josephson energies which changed compared to CD1 due to the aging of the junctions.

\section{Data Analysis} \label{section: data analysis}
\begin{figure}[h!]
\advance\leftskip-0.5cm
\includegraphics[width=\columnwidth]{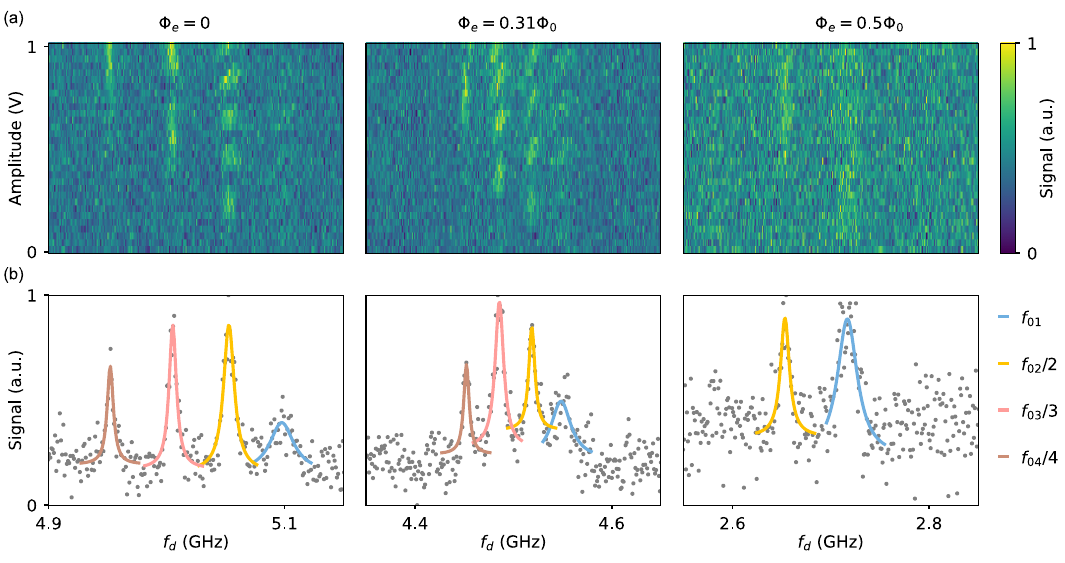} 
\caption{\label{fig_supp_analysis} Extraction of the transitions $f_{01}$, $f_{02} / 2$, $f_{03} / 3$ and $f_{04} / 4$ at $\Phi_\textrm{e} = 0$ , at $\Phi_\textrm{e} = 0.31\Phi_0$ which corresponds to $\lambda \approx 0.98$ and at $\Phi_\textrm{e} = \Phi_0/2$ (CD2). (a) Two-tone spectroscopy versus amplitude. (b) Raw data averaged over the amplitude to facilitate the fitting of the transitions to a Lorentzian function.}
\centering
\end{figure}

Figure \ref{fig_supp_analysis} presents the data analysis done in order to extract the transition frequencies from the raw data. We started from performing a two-tone power spectroscopy at each flux point, with the presented data given at $\Phi_\textrm{e} = 0$, $\Phi_\textrm{e} = 0.31\Phi_0$ which corresponds to $\lambda \approx 1$, and $\Phi_\textrm{e} = \Phi_0/2$. Next, the signal was averaged over the drive amplitude to reduce background noise. The amplitude range was adapted for each transition and flux point, which resulted in power broadening of the $f_{01}$ transition. Finally, we fit a Lorentzian to each one of the transitions and extracted the frequency. Near $\Phi_\textrm{e} = \Phi_0/2$  we were only able to measure two transitions, $f_{01}$ and $f_{02}/2$. 

\section{Full circuit analysis
\label{section_full_circuit}}

Here we perform a full circuit analysis of the double-junction transmon coupled to a resonator. The Lagrangian of our system is of the form,
\begin{equation}
    \mathcal{L} = T(\{\dot\Phi\}) - U(\{\Phi\}),
\end{equation}
where $T$ and $U$ are the kinetic and potential energies respectively \cite{terhal_lecture_notes}. We work in the $\Phi_1, \Phi_2, \Phi_\textrm{r}$-basis, where $\Phi_1,\Phi_2$ are the flux differences across the single junction and the SQUID respectively, corresponding to the reduced phase variables $\phi_i = 2\pi\Phi_i/\Phi_0$ for $i=1,2$, and $\Phi_r$ is the flux difference across the quarter-wave resonator which is modelled as a capacitor, $C_r$, in parallel with an inductor, $L$. 
The kinetic energy of the device in this basis can be written as,
\begin{equation}
    T = \frac{C}{2}(\dot{\Phi}_1 + \dot{\Phi}_2)^2 + \frac{C_{J1}}{2}(\dot{\Phi}_1)^2 + \frac{C_{J2}}{2}(\dot{\Phi}_2)^2 + \frac{C_r}{2}(\dot{\Phi}_r)^2 + \frac{C_g}{2}(\dot{\Phi}_r - \dot{\Phi}_1 - \dot{\Phi}_2)^2,
\end{equation}
where $C$ corresponds to the capacitance of the qubit island to ground, $C_{Ji}$ corresponds to the self-capacitance for the single junction and the SQUID respectively and $C_g$ to the coupling between qubit and the resonator. All capacitance parameters are understood to be effective values that include parasitic contributions from the circuit layout.
The potential energy term is given by,
\begin{equation}
    U(\Phi_1, \Phi_2,\Phi_\textrm{r}) = - E_{J1}\cos{\left(\frac{2\pi\Phi_1}{\Phi_0}\right)} - E_{J2}(\phi_\textrm{e})\cos{\left(\frac{2\pi\Phi_2}{\Phi_0}\right)} + \frac{1}{2L}\Phi_{\textrm{r}}^2,
\end{equation}
where $\phi_{e} = 2\pi\Phi_{e}/\Phi_0$ with $\Phi_{e}$ being the external flux applied through the SQUID loop.
We define the corresponding charge term for each of the flux terms:
\begin{equation}
    Q_1 = (C + C_{J1} + C_g)\dot\Phi_1 + (C + C_g)\dot\Phi_2 - C_g\dot\Phi_r,
\end{equation}
\begin{equation}
    Q_2 = (C + C_g)\dot\Phi_1 + (C + C_{J2} + C_g)\dot\Phi_2 - C_g\dot\Phi_r,
\end{equation}
\begin{equation}
    Q_r = (C_r + C_g)\dot\Phi_r - C_g(\dot\Phi_1 + \dot\Phi_2).
\end{equation}
Now, we can write the capacitance matrix of the device as, 
\[\mathbb{C} = 
\begin{pmatrix}
C + C_{J1} + C_g & C + C_g & -C_g \\
C + C_g & C + C_{J2} + C_g & -C_g \\
- C_g & -C_g & C_r + C_g
\end{pmatrix}.
\]
The Hamiltonian of the system can be derived using,
\begin{equation}
    H_{\textrm{full}} = \frac{1}{2}Q^T\mathbb{C}^{-1}Q + U(\Phi_1, \Phi_2, \Phi_\textrm{r}).
\end{equation}
Introducing the dimensionless charge and phase variables, $n = Q/(2e)$ and $\phi = 2\pi\Phi/\Phi_0$, respectively, we can rewrite the Hamiltonian as,
\begin{equation} \label{eq_supp_ham_full}
    H_{\textrm{full}} = 4E_{C1}n_1^2 + 4E_{C2}n_2^2 + 4E^r_{C}n_{\textrm{r}}^2 + g_{12}n_1n_2 + g_{1,\textrm{r}}n_1n_{\textrm{r}} + g_{2,\textrm{r}}n_2n_{\textrm{r}}  + U(\phi_1, \phi_2,\phi_\textrm{r})
\end{equation}
with 
\begin{equation}
    U(\phi_1, \phi_2,\phi_\textrm{r}) = - E_{J1}\cos{\phi_1} - E_{J2}(\phi_\textrm{e})\cos{\phi_2} + \frac{E_L}{2}\phi_{\textrm{r}}^2,
\end{equation}
where $E_L = \Phi_0^2 / (4\pi^2L)$ is the inductive energy of the resonator.

\begin{figure}
\advance\leftskip-0.5cm
\includegraphics[width=246pt]{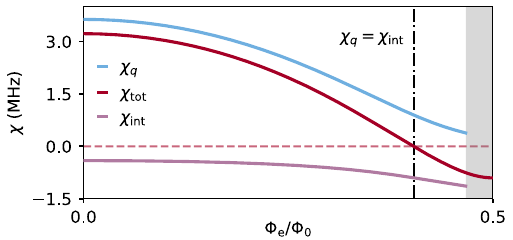} 
\caption{\label{fig_supp_disp_shift} Simulation of the dispersive shift, $\chi_0$, as a combination of the dispersive shift applied by the qubit mode and by the internal mode as a function of flux. The dotted-dashed line represents the flux value where the total dispersive shift is cancelled. The gray area represents the flux values where there is an avoided crossing between the $f_{01}$ of the internal mode and the $f_{03}$ of the qubit mode.}
\centering
\end{figure}

The charging energies and couplings are defined as follows,
\begin{equation}
    E_{Ci} = \frac{e^2}{2}\frac{C + C_{Jj}}{C'(C_1+C_2) + C_1C_2},
\end{equation}
\begin{equation}
    E^r_{C} = \frac{e^2}{2}\frac{C'' + C_g}{C''(C_g+C_r) + C_gC_r},
\end{equation}
\begin{equation}
    g_{i,r} = 4e^2\frac{C_{Jj}C_g}{(C_g+C_r)(C'(C_1+C_2) + C_1C_2)},
\end{equation}
\begin{equation}
    g_{12} = -4e^2\frac{C'}{C'(C_1+C_2) + C_1C_2},
\end{equation}
for $i,j = 1,2$ and $i \neq j$, with $C' = C + C_gC_r/(C_g+C_r)$ and $C'' = C + C_{J1}C_{J2}/(C_{J1}+C_{J2})$.

The full Hamiltonian makes it possible to study the interaction between the resonator and the double-junction transmon. Applying a Schrieffer-Wolff transform \cite{blais_rev} allows us to calculate the total dispersive shift $\chi_0 = \chi_q + \chi_\textrm{int}$ applied by the first excited states of the qubit mode, $\chi_q$, and of the internal mode, $\chi_\textrm{int}$. Each contribution can be calculated through,
\begin{equation}
    \chi_q =  \frac{hf_r}{16E^r_{C}}\frac{|g_{1,r}\mu_{10,1} + g_{2,r}\mu_{10,2}|^2}{\Delta_{10}},
\end{equation}
\begin{equation}
    \chi_\textrm{int} =  \frac{hf_r}{16E^r_{C}} \frac{|g_{1,r}\mu_{01,1} + g_{2,r}\mu_{01,2}|^2}{\Delta_{01}},
\end{equation}
where $\mu_{ij,k} = \braket{i_qj_\textrm{int}|n_k|0_q0_\textrm{int}}$ for $i,j = 0,1$, $k=1,2$, $hf_r = \sqrt{8E^r_{C}E_L}$ and $\Delta_{ij} = f_r-f_{ij}$.

\section{Born-Oppenheimer Approximation}
Here we review the Born-Oppenheimer (BO) approximation of the double-junction circuit (neglecting the coupling to the resonator) following Ref.~\cite{leo_paper}. In order to apply the BO approximation, we work in the qubit and internal-mode basis \cite{Rymarz2023}, with $\phi_q = \phi_1 + \phi_2$ and $\phi_\textrm{int} = (C_{J2}\phi_2 - C_{J1}\phi_1) / (C_{J1} + C_{J2})$. After plugging these definitions into the expression for the kinetic energy of the device we can rewrite the capacitance matrix in the new basis,
\[\mathbb{C_\textrm{q}^\textrm{int}} = 
\begin{pmatrix}
C + \frac{C_{J1}C_{J2}}{C_{J1}+C_{J2}} & 0 \\
0 & C_{J1} + C_{J2}
\end{pmatrix}.
\]

Now we can also write the Hamiltonian in the new basis where the coupling term between the two modes disappears.
\begin{equation}
    H(\phi_q,\phi_\textrm{int}) = 4E^q_{C}n_q^2 + 4E^\textrm{int}_{C}n_\textrm{int}^2 - E_{J\Sigma}\cos{\frac{\phi_q}{2}\cos{\left(\phi_\textrm{int} - k\frac{\phi_q}{2}\right)}} + E_{J\Delta}\sin{\frac{\phi_q}{2}\sin{\left(\phi_\textrm{int} - k\frac{\phi_q}{2}\right)}}
\end{equation}
where the charging energies associated with the qubit and internal modes are $E^q_C = e^2/2(C+C_{J1}C_{J2} / (C_{J1} + C_{J2})) $ and $E_C^\mathrm{int} = e^2/2(C_{J1}+C_{J2}) $ while $E_{J\Sigma}=E_{J1}+E_{J2}$, $E_{J\Delta}=E_{J1}-E_{J2}$ are the sum and differences of the Josephson energies with $k = (C_{J1} - C_{J2}) / ({C_{J1} + C_{J2}})$ as the junction capacitance asymmetry.

\begin{figure}
\advance\leftskip-0.5cm
\includegraphics[width=\columnwidth]{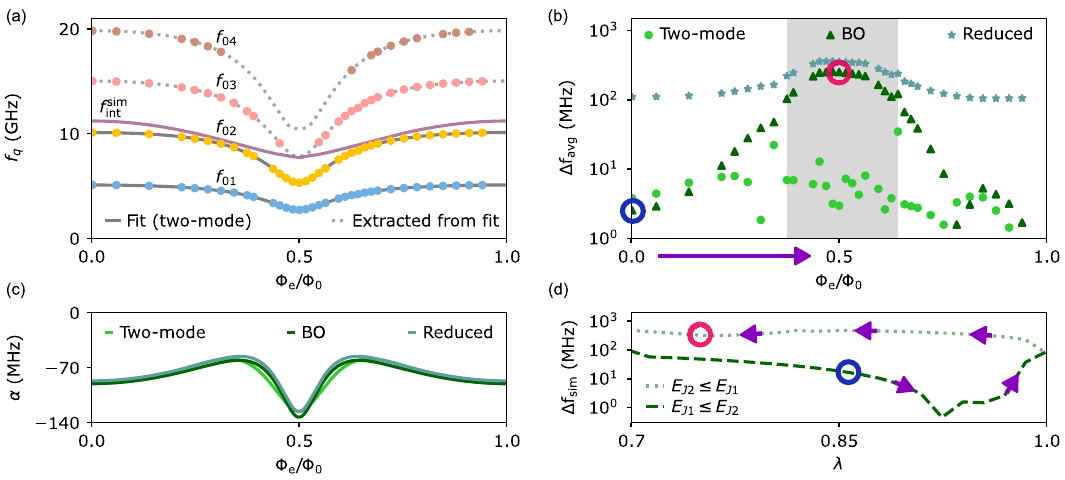} 
\caption{\label{fig_supp_full_fit} Measurement of the four transitions as a function of external flux (CD2) with the fit to the two-mode model and the fit error for each of the models. (a) Measurement of the four observed transitions which were extracted from the raw data, with the lowest two transitions being used to fit the data to the full two-mode model from Eq.~\ref{eq_supp_ham_full}. Simulation of the internal mode first excited state using the parameters extracted from the fit. (b) Average fit error of the two-mode, BO and reduced models, compared to the measured values. The shaded area corresponds to the region where the Josephson energy of the SQUID is smaller than the energy of the single junction resulting in a larger BO error. $\Phi_\textrm{e} = 0$ $(\Phi_\textrm{e} = 0.5\Phi_0)$ is marked with a blue (pink) circle. The purple arrow represents tuning of the flux from $\Phi_\textrm{e} = 0$ to $\Phi_\textrm{e} = 0.5\Phi_0$. (c) Anharmonicity as a function of flux using the extracted fit parameters in the two-mode, BO and reduced models. (d) Average simulated frequency discrepancy for the first four qubit transitions between the two-mode model and the BO model as a function of $\lambda$. The two branches correspond to the case when the Josephson energy of the single junction is smaller than that of a SQUID (dark green), therefore aligned with $C_{J1} < C_{J2}$, and the opposite (light green). At $\Phi_\textrm{e} = 0$ (blue circle) $E_{J1} < E_{J2}$ therefore it is on the aligned branch, and at $\Phi_\textrm{e} = 0$ (pink circle) $E_{J2} < E_{J1}$, therefore it is on the misaligned branch. The purple arrows represent the path followed by the device parameters as a function of external flux.}
\centering
\end{figure}

We can rewrite the potential energy of the two junctions in series using standard trigonometric identities,
\begin{align}\label{eq:U_prime}
    U'(\phi_q,\phi_\textrm{int})&=- s(\phi_q) E_{J\Sigma}\cos\left(\phi_\textrm{int}-f(\phi_q)\right)\sqrt{1-\lambda\sin^2\left(\phi_q/2\right)},\\
    f(\phi_q)&=k\frac{\phi_q}{2}-\arctan\left(\frac{E_{J\Delta}}{E_{J\Sigma}}\tan\frac{\phi_q}{2}\right),
\end{align}
where $s(\phi)=\mathrm{sign}[\cos(\phi/2)]$ is a sign.

To obtain the effective one-dimensional Hamiltonian, we apply a Born–Oppenheimer approximation that exploits the energy separation between the fast internal mode $\phi_\textrm{int}$ and the slow collective mode $\phi_q$. Treating $\phi_q$ as a classical parameter and assuming the internal mode remains in its ground state implies that the internal coordinate localizes at its minimum. As a result, any non-diverging function $f(\phi_q)$ yields the same effective single-mode BO potential $U_\mathrm{BO}(\phi_q)$, despite corresponding to different underlying two-mode potentials. This allows us to adopt a simplified form by choosing $f(\phi_q)=0$ and $s(\phi_q)=+1$, yielding
\begin{align}\label{eq:simple_potential}
U_\mathrm{simple}(\phi_q) = -E_{J\Sigma}\cos\phi_\mathrm{int}\sqrt{1-\lambda\sin^2\left(\phi_q/2\right)}.
\end{align}

\begin{figure}
\advance\leftskip-0.5cm
\includegraphics[width=\columnwidth]{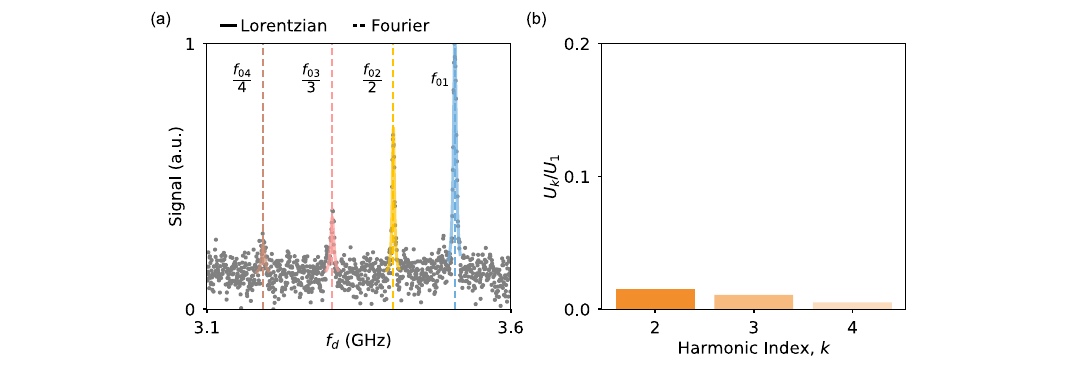} 
\caption{\label{fig_supp_FFT} Two-tone spectroscopy of a traditional single-junction transmon as a reference device, and its extracted harmonic content (CD2). (a) Two-tone spectroscopy showing four qubit transitions with Lorentzians around the measured transition frequency (solid lines) and the extracted frequencies (dashed lines) from the fit to the phenomenological model, $H_\textrm{T} = 4E_Cn^2 - \sum^{4}_{k=1}U_k\cos(k\phi)$. (b) The extracted harmonic content from the fitting to the phenomenological model.}
\centering
\end{figure}

We separate the simplified potential in Eq.~\eqref{eq:simple_potential} into slow and fast potentials to uncover the slow and fast Hamiltonians
\begin{align}
    H^q_\mathrm{slow} &=4E^q_Cn^2_q- E_{J\Sigma}\sqrt{1-\lambda\sin^2\left(\phi_q/2\right)},  \label{eq:H_slow_qubit}\\ 
    H^\mathrm{int}_\mathrm{fast} &=4E_C^\mathrm{int}n^2_\mathrm{int} + E_{J\Sigma}\sqrt{1-\lambda\sin^2\left(\phi_q/2\right)}\left( 1-\cos\phi_\mathrm{int} \right). \label{eq:H_fast_int}
\end{align}

In the regime $E_{J\Sigma}\gg E_C^\mathrm{int}$, the internal mode is localized near the origin, allowing the approximation $1-\cos\phi_\mathrm{int}\approx \phi_\mathrm{int}^2/2$, such that its harmonic ground-state energy yields the BO correction potential,
\begin{equation}\label{eq:U_corr}
    U_\mathrm{BO}(\phi_q)= E_{J\Sigma}\sqrt{\frac{2E_C^\mathrm{int}}{E_{J\Sigma}}\sqrt{1-\lambda\sin^2\left(\phi_q/2\right)}},
\end{equation}
and the total BO Hamiltonian thus becomes
\begin{align}\label{eq:H_BO_final}
    H_\mathrm{BO} &= 4E^q_Cn_q^2 - E_{J\Sigma}\sqrt{1-\lambda\sin^2\left(\phi_q/2\right)} + U_\mathrm{BO}(\phi_q).
\end{align}

Physically, the Born–Oppenheimer procedure integrates out the high-frequency internal oscillation, renormalizing the Josephson energy and introducing a small correction of order $(E_{C}^\textrm{int}/E_{J\Sigma})^{1/2}$ that captures the zero-point motion of the fast mode.

Figure~\ref{fig_supp_full_fit}a presents all four transitions ($f_{01}$, $f_{02} / 2$, $f_{03} / 3$, $f_{04} / 4$) that were measured as a function of flux. The lowest two transitions ($f_{01}$, $f_{02} / 2$) were fit using the full two-mode model described in circuit analysis section, which provided a good prediction for the $f_{03}$, $f_{04}$ transitions (dotted gray lines). Using the parameters extracted from the fit we simulated the first excited state of the internal mode (purple).

The average discrepancy between the extracted transition frequencies from the Lorentzian fitting of the data and the two-mode, Born-Oppenheimer (BO) and reduced models can be seen in Fig.~\ref{fig_supp_full_fit}b. We observe a good fitting to the full two-mode model across all flux values with the average difference in frequency being $\sim 6 \, \si{\mega\hertz}$ and with only two flux points having a difference of above $\sim 20 \, \si{\mega\hertz}$. For the BO model we observe an equally good fit as for the two-mode model next to $\Phi_\textrm{e} = 0$, but closer to $\Phi_\textrm{e} = \Phi_0/2$ the model diverges up to $\sim200\si{\mega\hertz}$ compared to the measured values. Despite the differing levels of agreement with the measured transition frequencies shown in Fig.~\ref{fig_supp_full_fit}b, the flux dependence of the anharmonicity is nearly identical for all three models, as can be seen in Fig.~\ref{fig_supp_full_fit}c. This is because the $f_{01}$ and $f_{02}/2$ transitions are shifted by approximately the same amount in both the BO and reduced models. The discrepancy between the anharmonicity extracted from the reduced model in Figure 2b in the main text and in Fig.~\ref{fig_supp_full_fit}c arises from the use of the analytical expression in the former and the numerical model in the latter.

To investigate the regime of validity of the BO approximation, we simulate the discrepancy between the two-mode and BO models as a function of $\lambda$ for two junctions in series while keeping the charging energies fixed (Fig.~\ref{fig_supp_full_fit}d). The Josephson energy of one junction is held constant, and for each value of $\lambda$ the energy of the second junction is adjusted accordingly. The accuracy of the BO approximation is governed by the interplay between capacitance and junction energy asymmetries: for small $\phi_q$, Eq.~\ref{eq:U_prime} yields $f(\phi_q)\approx (k - E_{J\Delta}/E_{J\Sigma})\phi_q/2$, where $k=(C_{J1}-C_{J2})/(C_{J1}+C_{J2})$ is fixed by the capacitances, while $E_{J\Delta}/E_{J\Sigma}$ captures the flux-tunable junction energy asymmetry. The approximation performs well when these contributions are comparable in magnitude and sign, and degrades otherwise. This yields two distinct regimes: one where the first junction has lower energy than the second (aligned with $C_{J1} < C_{J2}$, see Table~\ref{tab:parameter_tab}), and one where it has higher energy. We find that the discrepancy between the BO and two-mode models differs by one to three orders of magnitude between these regimes. Furthermore, within the lower-error branch (dark green dashed line in Fig.~\ref{fig_supp_full_fit}d), varying $\lambda$ leads to changes in the discrepancy of up to two orders of magnitude.

\begin{figure}
\advance\leftskip-0.5cm
\includegraphics[width=\columnwidth]{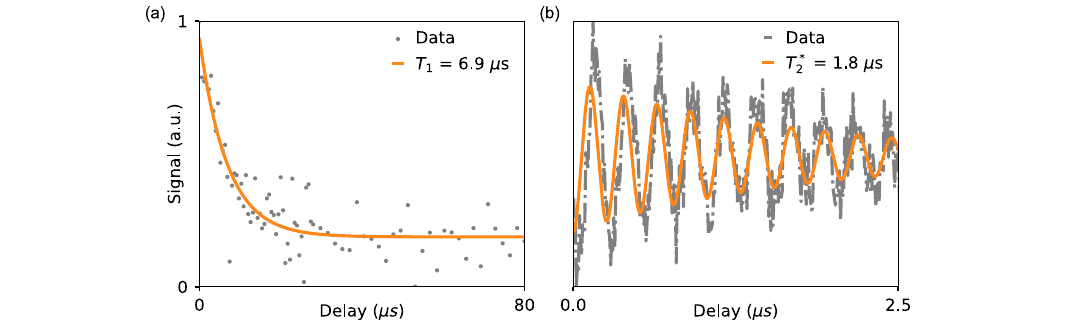} 
\caption{\label{fig_supp_coherence}  Coherence data at $\Phi_\textrm{e} = 0$ (CD1). (a) Qubit relaxation time measurement which was fit to be $T_1 = 6.9 \, \si{\micro\s}$. (b) Ramsey interferometry measurement. The qubit dephasing time was fit to be $T^*_2 = 1.8 \, \si{\micro\s}$.}
\centering
\end{figure}

This behavior is consistent with Fig.~\ref{fig_supp_full_fit}b: near $\Phi_\textrm{e} = 0$ (blue circle), the device operates in the low-error regime with $\lambda \approx 0.86$, resulting in discrepancies on the order of $10,\si{\mega\hertz}$. As the flux is tuned, the Josephson energy of the SQUID loop is modified, and the system moves out of this low-discrepancy region, leading to a growing BO error (purple arrows in Fig.~\ref{fig_supp_full_fit}d). When the SQUID Josephson energy becomes smaller than that of the single junction (shaded region in Fig.~\ref{fig_supp_full_fit}b), the system transitions to the high-error branch (light green dotted line in Fig.~\ref{fig_supp_full_fit}d) which is observed especially near $\Phi_\textrm{e} = 0.5\Phi_0$ (pink circle). This also explains why identical values of $\lambda$ can correspond to different discrepancy values. This dependence can be understood from Eq.~\ref{eq:U_prime}: for small $\phi_q$, $f(\phi_q)\approx (k - E_{J\Delta}/E_{J\Sigma})\phi_q/2$, where $k=(C_{J1}-C_{J2})/(C_{J1}+C_{J2})$ is fixed by the capacitances, while $E_{J\Delta}/E_{J\Sigma}$ captures the junction energy asymmetry which is tunable and changes its sign near half flux quantum. The BO approximation performs well when these two contributions are comparable in magnitude and sign, and degrades once they differ significantly.

\section{Fixed-frequency Transmon Data}

To provide a comparison for the expected harmonic content of an SIS junction, we fabricated and measured a fixed-frequency transmon observing four transitions (Fig.~\ref{fig_supp_FFT}a). To extract the harmonic content, we fit the spectra to a phenomenological model, $H_\mathrm{T} = 4E_C n^2 - \sum_{k=1}^{4} U_k \cos(k\phi)$, and extract the corresponding transitions (dashed lines). The resulting harmonics (Fig.~\ref{fig_supp_FFT}b) show weak higher-order contributions, $U_k = [1, 0.015, 0.011, 0.005]$, consistent with previously reported values for SIS junctions.

\section{Coherence}

Figure~\ref{fig_supp_coherence} shows coherence data measured at $\Phi_\textrm{e} = 0$. This data is representative of a series of coherence measurements done at different flux points which resulted in the qubit relaxation time, $T_1$, between $4 \, \si{\micro\s}$ and $13 \, \si{\micro\s}$. Due to increasing flux noise it was challenging to perform an adequate fitting of the qubit dephasing time away from the flux sweet spots. Fig.~\ref{fig_supp_coherence}b shows a characteristic Ramsey interferometry measurement resulting in $T^*_2\sim1.8\, \si{\micro\s}$.

\input{references_SI.bbl}

%% file: preamble.tex
\documentclass[reprint, superscriptaddress, amsmath,amssymb,
 aps]{revtex4-2}

\usepackage{siunitx}
\usepackage{graphicx}
\usepackage{dcolumn}
\usepackage{bm}
\usepackage{braket}
\usepackage{comment}
\usepackage{hyperref}
\usepackage{cleveref}
\usepackage{subfiles}
\usepackage{tikz}
\newcommand{\jj}{%
  \tikz[baseline=-0.6ex, x=1em, y=1em, yshift=-0.7pt]{%
    \def\s{0.18}   
    \def\sep{0.6} 
    \def\lead{0.35}

    \draw (-\lead,0) -- (-\s,0);

    \draw (-\s,-\s) rectangle (\s,\s);
    \draw (-\s,-\s) -- (\s,\s);
    \draw (-\s,\s) -- (\s,-\s);

    \draw (\s,0) -- (\sep-\s,0);

    \begin{scope}[xshift=\sep em]
      \draw (-\s,-\s) rectangle (\s,\s);
      \draw (-\s,-\s) -- (\s,\s);
      \draw (-\s,\s) -- (\s,-\s);
    \end{scope}

    \draw (\sep+\s,0) -- (\sep+\lead,0);
  }%
}